\documentclass[
    bibnotes,
    amsmath,amssymb,amsfonts,
    aps,
    a4paper,
]{quantumarticle}

\pdfoutput=1

\usepackage[a4paper, total={6in, 8in}]{geometry}
\usepackage{graphicx}
\usepackage{dcolumn}
\usepackage{bm}
\usepackage{hyperref}
\usepackage{braket}
\usepackage{dsfont}
\usepackage{color}
\usepackage{soul}
\usepackage[caption=false]{subfig}

\newcommand{\be}{\begin{equation}}
\newcommand{\ee}{\end{equation}}
\newcommand{\ba}{\begin{array}}
\newcommand{\ea}{\end{array}}
\newcommand{\bqa}{\begin{eqnarray}}
\newcommand{\eqa}{\end{eqnarray}}

\begin{document}


\title{Analogue Quantum Simulation with Fixed-Frequency Transmon Qubits}

\author{Sean Greenaway}
\affiliation{Physics Department, Blackett Laboratory, Imperial College London, Prince Consort Road, SW7 2BW, United Kingdom}
\author{Adam Smith}
\affiliation{School of Physics and Astronomy, University of Nottingham, Nottingham, NG7 2RD, UK}
\affiliation{Centre for the Mathematics and Theoretical Physics of Quantum Non-Equilibrium Systems, University of Nottingham, Nottingham, NG7 2RD, UK}
\author{Florian Mintert}
\affiliation{Physics Department, Blackett Laboratory, Imperial College London, Prince Consort Road, SW7 2BW, United Kingdom}
\affiliation{Helmholtz-Zentrum Dresden-Rossendorf, Bautzner Landstraße 400, 01328 Dresden, Germany}
\author{Daniel Malz}
\affiliation{Max-Planck-Institute of Quantum Optics, Hans-Kopfermann-Str. 1, 85748 Garching, Germany}
\affiliation{Department of Physics, Technische Universität München, James-Franck-Straße 1, 85748 Garching, Germany}

\date{23rd January 2024}

\begin{abstract}
We experimentally assess the suitability of transmon qubits with fixed frequencies and fixed interactions for the realization of analogue quantum simulations of spin systems. We test a set of necessary criteria for this goal on a commercial quantum processor using full quantum process tomography and more efficient Hamiltonian tomography. Significant single qubit errors at low amplitudes are identified as a limiting factor preventing the realization of analogue simulations on currently available devices. We additionally find spurious dynamics in the absence of drive pulses, which we identify with coherent coupling between the qubit and a low dimensional environment. With moderate improvements, analogue simulation of a rich family of time-dependent many-body spin Hamiltonians may be possible.
\end{abstract}

\maketitle

\section{Introduction}
Recent experimental progress towards the development of fault-tolerant quantum computers has been considerable~\cite{arute2019quantum}. However, the current so-called {\it noisy intermediate scale quantum} (NISQ) devices are limited by a level of noise that at present precludes implementation of many algorithms~\cite{preskill2018quantum}. An exciting application which is thought to be achievable even in the presence of noise, lies in the quantum simulation of physical systems for which classical simulations are intractable~\cite{georgescu2014quantum}.

A wide array of experimental platforms have already demonstrated many of the commonly-applied criteria~\cite{cirac2012goals} for the realization of quantum simulations~\cite{blais2020quantum, blatt2012quantum, browaeys2020many,lloyd1996universal}. The digital approach towards implementing such a simulation typically involves decomposing the time evolution operator into a series of implementable gates through Trotterization~\cite{trotter1959product}, an approach that has been demonstrated for a variety of small systems experimentally on NISQ devices~\cite{lanyon2011universal,barends2016digitized,peng2005quantum}. Such gate-based quantum simulations are highly flexible, being capable (in principle at least) of simulating any quantum system due to the universality of quantum computation. In practise, these simulations are restricted to small system sizes and short simulation times, since increasing either necessitates more gates, which come with a commensurate increase in error.

An alternative approach, known as {\it analogue quantum simulation}~\cite{georgescu2014quantum} directly simulates a system of interest by manipulating a controllable experimental system that mimics it, allowing for decomposition protocols such as Trotterization to be circumvented. The increased efficiency of analogue simulation has allowed for the simulation of larger quantum systems for longer times than gate-based approached in platforms such as cold atoms~\cite{endres2011observation,greif2013short,greiner2002quantum} and has motivated substantial research into implementations in other platforms such as superconducting circuits~\cite{houck2012chip,hartmann2016quantum,wilkinson2020superconducting,hartmann2008quantum,roushan2017spectroscopic,zhu2022observation}. This efficiency comes at the expense of limiting the simulations to the system's ``native" Hamiltonians. Additionally, such analogue simulators may be restricted in the measurements that can be performed upon them, in contrast to gate-based devices which have access to general Pauli string measurements. This motivates the search for alternative quantum simulation platforms to complement the existing ones.

A typical workflow for implementing analogue quantum simulations is to choose a target Hamiltonian of interest and then to construct a control protocol that maps the system Hamiltonian to that target. Thus, in order to implement an analogue quantum simulation, it is crucial that the map between an applied control protocol and the resulting experimental effective Hamiltonian is well understood. To this end three criteria may be identified that are necessary for the experimental implementation of analogue quantum simulations~\cite{cirac2012goals,divincenzo2000physical}:
\begin{itemize}
    \item[(C1)] {\it Expressibility:} The experimental Hamiltonian must permit control protocols which allow for some class (or multiple classes) of interesting models to be simulated.
    \item[(C2)] {\it Practical Controllability:} It should be possible to switch individual control terms on and off independently, without inducing significant errors on other qubits.
    \item[(C3)] {\it Stability:} The map between the control protocol and the experimental effective Hamiltonian should be stable enough over time that characterization and simulation experiments can be performed without the map changing due to, for example, parameter drift. Additionally, the coherence time of the device should be sufficiently long to allow simulations to be performed.
\end{itemize}In this work we experimentally assess the extent to which fixed-frequency, fixed-interaction (FF) transmon qubits available through the IBM Quantum cloud-based platform~\cite{Qiskit2} satisfy these criteria, and thereby probe the utility of this system as a platform for analogue quantum simulation. As a platform primarily used for gate-based quantum computation, FF transmon qubits allow for the control of individual qubits and arbitrary Pauli string measurements~\cite{krantz2019quantum}. Additionally, the underlying physical Hamiltonian used to implement these gates may be mapped to a wide array of interesting Hamiltonians which may be simulated~\cite{malz2021topological}. As such, FF transmon devices are potentially highly useful as a platform for analogue simulation.

FF transmon devices can be modelled as weakly coupled Duffing oscillators controlled via time-dependent drive pulses. Two-body entanglement can be generated through a cross-resonance interaction, in which a qubit is driven at the resonant frequency of another to which it is coupled~\cite{rigetti2010fully,chow2011simple}. This procedure results in an entangling operation comprised of $ZX$ and $ZY$ terms (where $\mathds{1},X,Y,Z$ are the Pauli matrices and the tensor product is implied) along with a number of spurious single qubit terms. In order to use this platform for analogue quantum simulation, the magnitude of these terms must be well known such that they can be controlled during a simulation. If this can be achieved, FF transmon qubits should allow for a rich class of systems to be simulated, including Ising Hamiltonians with individually addressable Ising coupling and single qubit magnetic field control, systems with $XY$-type interactions and the quantum East model~\cite{pancotti2020quantum}.

While the effective Hamiltonian resulting from applying cross-resonance drives can be derived rigorously using Floquet theory~\cite{malekakhlagh2020first}, the resulting predictions are not sufficiently precise to run high-fidelity simulations. Instead, we characterize the device experimentally. While the aim of the paper is to assess the platform against the criteria listed above, each experiment yields evidence about multiple criteria. For clarity, the experimental results are presented in full first, with a summary of the assessments of the above criteria presented at the end of the paper.

In the first set of experiments, we use full quantum process tomography (QPT) to show that it is possible to individually and independently control the cross-resonance interaction, and to find the dominant spurious terms. Second, we use Hamiltonian tomography to accurately and efficiently extract the Hamiltonian rates.

We characterize the unwanted single-qubit terms generated through the cross-resonance drives and extract phase, amplitude and detuning errors. In principle, these can be cancelled using weak additional tones, but we observe that the drive amplitudes cannot be controlled with sufficient precision to do so, which could be fixed with hardware improvements.

Additionally, we identify spurious dynamics in the absence of driving with coupling between two level system (TLS) defects and the transmon qubits. The error terms arising from these fluctuate significantly over time. This, rather than imperfections with the entangling operation, is identified as the key limiting factor preventing the realization of fully-controllable analogue quantum simulations on current-generation FF transmon devices.

\section{Controlling fixed-frequency, fixed-interaction transmon qubits}\label{sec:background}
The starting point for the analysis presented here is the verification of the model used to inform experimental control protocols. For FF transmon qubits, the system can be described as a series of $n$ coupled Duffing oscillators, for which the Hamiltonian is~\cite{krantz2019quantum, malz2021topological}
\begin{align}
    H^{\text{duff}} &= \sum_{i=1}^n\left(\omega_i a_i^\dagger a_i + \alpha_i a_i^\dagger a^\dagger_i a_i a_i + D_i(t)(a_i + a_i^\dagger)\right) \nonumber \\
    &+ \sum_{\langle i,j \rangle} J_{ij}(a_i - a_i^\dagger)(a_j - a_j^\dagger) \ ,
\end{align}
where $\omega_i$ and $\alpha$ are the harmonic frequency and anharmonicity of the $i$th transmon respectively, $J_{ij}$ is the capacitive coupling strength between the $i$th and $j$th transmon, where the nearest-neighbour notation $\langle i,j \rangle$ reflects the physical connectivity of the device, and the drive on the $i$th transmon is given by
\begin{equation}
    D_i(t) = \frac{\Omega}{2}\operatorname{Re}\left[e^{i(\omega_i + \Delta_i)t}d_i(t)\right] \ ,
\end{equation}
with drive strength $\Omega$, applied detuning $\Delta_i$ and dimensionless drive envelope $d(t)$. For high enough anharmonicities relative to the applied drive strength, the transition between the states $\ket{0}$ and $\ket{1}$ is well separated from the higher energy levels, and so the system may be described by a qubit model,
\begin{equation}
    H = \sum_{i=1}^n \frac{\omega_i}{2}Z_i + D_i(t)X_i + \sum_{\langle i,j \rangle} J_{ij}Y_i Y_j \ ,
\end{equation}
where the notation $X,Y,Z$ has been used for the Pauli matrices.

Single qubit $X,Y$ and $Z$ terms can be independently controlled by applying a pulse at zero detuning with the drive envelope parameterized as
\begin{equation}\label{eq:single_drive_env}
    d_i(t) = (h^X_i(t) + ih^Y_i(t))\exp\left(-2i\Omega\int_0^{t}h^Z_i(t')dt' \right) \ .
\end{equation}
In the frame rotating at the qubit frequencies, the effective Hamiltonian resulting from this drive is
\begin{equation}\label{eq:single_qubit_eff}
    H(t)=\frac{\Omega}{2}\sum_i\left[h^X_i(t)X_i + h^Y_i(t)Y_i + h^Z_i(t)Z_i\right] \ .
\end{equation}
The coupling parameter $J_{ij}$ in FF transmon devices cannot be controlled, and thus needs to be small enough compared to the detuning between connected qubits that in the absence of driving the qubits are effectively decoupled. In this case, an entangling operation can be switched on by driving one qubit at the resonant frequency of another to which it is coupled~\cite{rigetti2010fully,sheldon2016procedure}. To first order, the off-resonant drive generates no dynamics. However, to second order interplay between the drive and the static coupling results in an effective {\it cross-resonance} entangling operation of the general form
\begin{equation}\label{eq:cr_ham_all_terms}
    H^{\text{CR}}_{ij} = \sum_{A\in\{\mathds{1}, X,Y,Z\}} c^{\mathds{1}A}_{ij}\mathds{1}_iA_j + c^{ZA}_{ij}Z_iA_j \ .
\end{equation}
Estimations for the values of the coefficients $\{c^k_{ij}\}$ have been extracted using high order Schrieffer-Wolff perturbation theory~\cite{malekakhlagh2020first,tripathi2019operation} but these depend on experimental parameters that are inaccessible to end users and which can drift over time. As a result, we find it more practical to extract the coefficients experimentally in a calibration process.

In the following sections, the results of evaluating and performing such an experimental calibration protocol are presented. To simplify the results such that the fundamental properties of the underlying system can be clearly identified, the phase of the drive pulse was calibrated such that the resulting dynamics should induce interactions along the $Z_iX_j$ axis only, and alternating qubits were driven such that the effective Hamiltonian should be of the form of an Ising-like system. This is not a fundamental limitation of the system, and the procedures described in this work should be generalizable such that the full range of dynamics expressible from Eq.~\eqref{eq:cr_ham_all_terms} can be generated.

\section{Hamiltonian Verification and Calibration Through Tomography}
The level of control offered by the cross-resonance Hamiltonian (Eq.~\eqref{eq:cr_ham_all_terms}) theoretically allows for the simulation of a rich array of lattice spin systems. However, it is crucial to verify that the experimental dynamics continue to satisfy criteria C1-C3 in the presence of imperfections such as cavity leakage. In particular, the cross-resonance drives should not induce any dynamics on qubits other than the control and target, and the result of applying two simultaneous cross-resonance drives on two different qubits should be the sum of the individual cross-resonance interactions (i.e., $H^{\text{eff}}_{1,2,3}=H^{\text{eff}}_{1,2} + H^{\text{eff}}_{2,3}$). Quantum process tomography (QPT) is ideally suited to the verification of these properties, since it makes no assumptions about the underlying dynamics and can thus identify unexpected terms. The experimental implementation of QPT for this purpose is presented in Sec.~\ref{sec:qpt}.

While QPT is rigorous, it is highly inefficient, requiring a number of expectation value measurements that is proportional to $d^4$ (with $d=2^n$ the dimension of the Hilbert space) to obtain the process matrix that characterizes a given channel. It is thus unsuitable for the calibration of the Hamiltonian coefficients in Eq.~\eqref{eq:cr_ham_all_terms}. Instead, having verified the structure of the effective Hamiltonian, these rates can be calibrated far more efficiently using {\it Hamiltonian tomography}~\cite{wang2015hamiltonian,li2020hamiltonian,ma2017hamiltonian}, which for this system requires only five Rabi oscillation experiments for a full characterization of the effective Hamiltonian. The Hamiltonian tomography protocol used in this work is presented in Sec.~\ref{sec:ham_tomog}.

\subsection{Quantum Process Tomography}\label{sec:qpt}
QPT yields the full process matrix characterizing the experimental dynamics, including non-unitary contributions from decoherence. For the purposes of Hamiltonian identification, only the unitary contribution of the full quantum channel is considered, and so these effects are not taken into account. The extent to which the dynamics are faithfully represented by a unitary operator is characterized by the dominant eigenvalue $\lambda_0$ of the process matrix, which is 1 for a fully unitary operator and $0<\lambda_0<1$ for non-unitary operations. Since non-unitary errors cannot be directly counteracted in the scheme proposed here, this measurement represents an additional check on the viability of analogue simulation protocols.

The effective Hamiltonian may be extracted by taking the logarithm of the reconstructed unitary evolution operator. Using this method, the eigenvalues of the effective Hamiltonian are obtained only up to factors of $2\pi$. This ambiguity can be alleviated by extracting Hamiltonian terms over a range of different pulse durations. 
\begin{figure}
    \includegraphics[width=\linewidth]{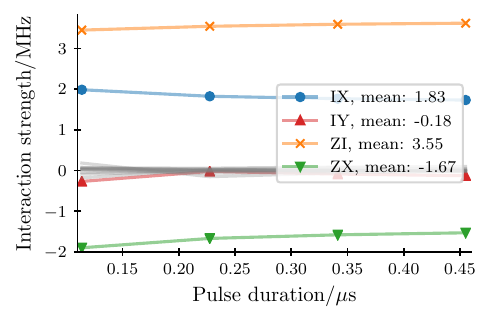}
    \caption{Effective Hamiltonian rates extracted from full process tomography on qubits 1 and 2, implemented on the \texttt{ibmq\_guadalupe} quantum device following the application of a cross-resonance drive on qubit 1 at the frequency of qubit 2 for varying pulse durations. The dominant $Z_iX_j$, $\mathds{1}_iX_j$ and $Z_i\mathds{1}_j$ terms are significantly stronger than any other Hamiltonian terms (shown in gray, with $\mathds{1}_iY_j$ shown in red as an illustration) and the rates are consistent for all drive durations as expected.}
    \label{fig:QPT_results_1}
\end{figure}
While the cross-resonance interaction can generate interactions consisting of any coherent mixture of $Z_iX_j$ and $Z_iY_j$ terms, for convenience here we work with interactions that are aligned along one axis only. This can be achieved by applying a drive pulse with a purely real drive envelope (i.e., $h^Y_i=0$ in Eq.~\eqref{eq:single_drive_env}), resulting in only $Z_iX_j$ interactions, along with the spurious single qubit rotations to be characterized. In practise, we find that the experimental pulse envelopes can accumulate significant phase errors. These can be eliminated by adjusting the phase of the drive envelope until no $Z_iY_j$ interactions are observed. This calibration procedure has been performed prior to performing all of the experiments presented here.

Fig.~\ref{fig:QPT_results_1} shows the observed effective Hamiltonian rates for a series of cross-resonance interactions applied on the \texttt{ibmq\_guadalupe} quantum device, with the minimum factors of $\pi$ needed to generate linear plots added. Factors of $\pi$ rather than $2\pi$ are added since the terms that dominate the dynamics ($Z_iX_j$, $\mathds{1}_iX_j$ and $Z_i\mathds{1}_j$) mutually commute, meaning that there is an additional ambiguity arising from the fact that adding $\pi$ to any term results in an unmeasureable global phase. These largest terms correspond to the terms predicted by Eq.~\eqref{eq:cr_ham_all_terms}, showing that the qualitative features of the cross-resonance gate are indeed reliable for this system. Additionally, the largest eigenvalue for the process matrices generated by these experiments was approximately $0.93$, showing that the dynamics are dominated by the unitary evolution. Given that the reported measurement error rates for IBM Quantum devices are on the order of $1\%$~\cite{sheldon2016procedure}, it is reasonable to ascribe a large portion of the non-unitary errors to state preparation and measurement errors.

The strengths of the dominant interactions and their magnitude relative to the subdominant terms are essentially constant over the measured pulse durations, indicating that the interactions in the system are stable with respect to the pulse duration. The applied drive pulse used to obtain these results had a strength of approximately $45 \text{MHz}$, which is well below the qubit-qubit coupling strength between the qubits considered (approximately $1.7 \text{MHz}$). In this weak driving regime, the system should be well approximated by the static cross-resonance Hamiltonian, which indeed seems to be the case. Any drive strength in this regime should give rise to $Z_iX_j$ and $\mathds{1}_1X_j$ terms that have similar relative magnitudes -- the $Z_i\mathds{1}_j$ term will have a different relative magnitude since it depends quadratically rather than linearly on the drive strength. In the strong driving regime, coupling to higher energy levels can induce substantial unwanted terms in the Hamiltonian. Since such a regime is difficult to control accurately enough for the purposes analogue quantum simulation, we do not consider it in this work and instead focus exclusively on the weak driving regime.

\begin{figure}
    \includegraphics[width=0.8\linewidth]{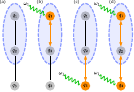}
    \caption{Schematic diagram for the series of process tomography experiments used to verify the qualitative form of the cross-resonance effective Hamiltonian Eq.~\eqref{eq:cr_ham_all_terms}. In all cases, full QPT is performed on qubits 1 and 2. (a) No drives are applied (b) Qubit 1 is driven at the frequency of qubit 2 (c) Qubit 3 is driven at the frequency of qubit 2 (d) Both qubit 1 and qubit 3 are simultaneously driven at the frequency of qubit 2.}
    \label{fig:QPT_schematic}
\end{figure}
To verify that the effective Hamiltonian for the three qubit channel is the sum of the two qubit effective Hamiltonians (i.e., $H^{\text{eff}}_{1,2,3}=H^{\text{eff}}_{1,2} + H^{\text{eff}}_{2,3}$), we use four two-qubit QPT experiments, where the tomography is performed on qubits 1 and 2 after the application of the following different drive protocols (outlined schematically in Fig.~\ref{fig:QPT_schematic}):
\begin{itemize}
    \item[1.] No drives, leave the qubits idle for the drive duration.
    \item[2.] Drive qubit 1, leave qubit 3 idle.
    \item[3.] Drive qubit 3, leave qubit 1 idle.
    \item[4.] Drive both qubit 1 and 3 simultaneously.
\end{itemize}
If the effective Hamiltonian generating the dynamics observed in experiment 4 is equal to the sum of Hamiltonians for the previous three experiments, then it can be concluded that the Hamiltonian for the full $n-$qubit system may be obtained by characterizing all qubit pairs involved in the experiment. Additionally, experiment 3 may be used to confirm that no unexpected additional terms are generated on idle qubits -- that is, the dynamics observed in experiment 3 should only be single qubit $X$ rotations.
\begin{figure}
    \includegraphics[width=\linewidth]{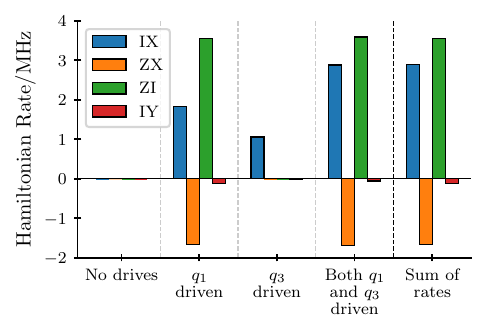}
    \caption{Summary of Hamiltonian rates obtained from the application of experiments 1-4 explained in the main text and summarised in Fig.~\ref{fig:QPT_schematic}, with the tomography performed on qubits 1 and 2 (shown by the blue ellipses), implemented on the \texttt{ibmq\_guadalupe} quantum device. Terms with rates less than 0.2 have been dropped for clarity, with the exception of $c_{\mathds{1}Y}$, which is shown for comparison. The first four sets of bars correspond to the experimental data for the various driving setups, while the final set corresponds to the predicted values for driving qubits 1 and 3 simultaneously, obtained as the sum of the Hamiltonian rates for driving qubits 1 and 3 separately plus the idle Hamiltonian, which is approximately vanishing. This predicted set of rates is very close (within $1.5\%$) to the real experimental rates for the simultaneous drive with the exception of the $\mathds{1}_iY_j$ rate, which is at a much lower magnitude and is therefore more susceptible to measurement error. This indicates that the simultaneous drive can indeed be modelled accurately as the sum of the individual effective Hamiltonians.}
    \label{fig:full_QPT_rates_diff_drives}
\end{figure}
Fig.~\ref{fig:full_QPT_rates_diff_drives} summarises the experimental results from experiments 1-4. As expected, no additional terms are observed on qubits 1 and 2 when qubit 3 is driven, regardless of whether qubit 1 is itself driven. Additionally, the predicted rates arising from adding the Hamiltonian rates for experiments 1-3 (final set of bars in Fig.~\ref{fig:full_QPT_rates_diff_drives}) matches the results of experiment 4 very well, with all the rates apart from $c_{\mathds{1}Y}$ (the rate of the $\mathds{1}_iY_j$ term) differing by less than $0.05$MHz, or less than $1.5\%$ of the observed Hamiltonian rates. The error for $c_{\mathds{1}Y}$ was slightly higher, at $0.07$MHz, however this is likely due to its significantly lower relative size, making it more susceptible to fluctuations due to measurement error. This indicates that the experimental drive behaves as indicated by Eq.~\eqref{eq:cr_ham_all_terms}, without significant cross-talk, and that the effective Hamiltonians for smaller subsystems can be added to obtain the dynamics for larger systems.

We note that we have not investigated longer-distance coupling on the basis that we have not observed cross-talk between next-nearest neighbours, which makes longer-range couplings unlikely. Under this assumption, the characterization of a full $n-$qubit system can be performed using a number of experiments that is independent of $n$ by characterizing disconnected sets of qubits in parallel.

\subsubsection{Three Qubit Quantum Process Tomography}
\begin{figure}
    \includegraphics[width=\linewidth]{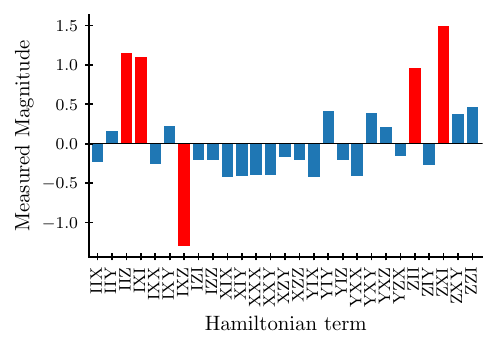}
    \caption{Hamiltonian terms resulting from applying simultaneous cross-resonance drives on qubits 1 and 3, implemented on the \texttt{ibmq\_guadalupe} quantum device. The Hamiltonian was obtained as the logarithm of the process matrix obtained from full process fidelity as described in the main text. The terms highlighted in red correspond to the expected non-zero coefficients arising from the cross-resonance drive and terms with magnitudes less than 0.15 have been dropped for clarity. While the results are significantly noisier than the two qubit experiments, the dominant terms are those expected from the theory.}
    \label{fig:3q_QPT_results}
\end{figure}
As an additional check that next-nearest-neighbour cross-talk is negligible, the full process matrix corresponding to the simultaneous cross-resonance drive on qubits 1 and 3 (that is, the same set of drives that generated the third set of data in Fig.~\ref{fig:full_QPT_rates_diff_drives}) can be evaluated to obtain the full three qubit Hamiltonian coefficients -- the results of such a procedure are shown in Fig.~\ref{fig:3q_QPT_results}. The primary reason for investigating the same drive protocol using three qubit QPT is to verify that the terms observed in Fig.~\ref{fig:full_QPT_rates_diff_drives} arise only from the expected two-body interactions and spurious single qubit terms and not from any unanticipated three-body interactions (such as a $Z_iX_jZ_k$ interaction). Although these terms should not be present based on the Hamiltonian Eq.\eqref{eq:cr_ham_all_terms}, it is important to verify that no unanticipated sources of cross-talk are present.

The dominant Hamiltonian terms in Fig.~\ref{fig:3q_QPT_results} are those expected from the theoretical drives (that is, $\mathds{1}_i\mathds{1}_jZ_k, \mathds{1}_iX_j\mathds{1}_k, \mathds{1}_iX_jZ_k, Z_i\mathds{1}_j\mathds{1}_k$ and $Z_iX_j\mathds{1}_k$), which are highlighted in red. There are a number of other terms with moderate strengths, albeit less than half the magnitude of the expected terms. Notably, these terms are completely unexpected, with no clear mechanism for how they could arise, whereas terms which could feasibly be generated in a system with substantial pulse leakage (such as $Z_iX_jZ_k$ or $Z_i\mathds{1}_jZ_k
$) are not observed. Moreover, there is no evidence of these terms in the two qubit Hamiltonians (Fig.~\ref{fig:full_QPT_rates_diff_drives}), which can be seen most clearly by the $Z_iZ_j\mathds{1}_k$ term, which has an observed strength approximately $1/3$ that of the $Z_iX_j\mathds{1}_k$ term in the three qubit QPT results, but which is negligible in the two qubit QPT results. This strongly implies that the observed terms are artefacts arising from the imperfect reconstruction of the Hamiltonian from the experimental data. Measurement error and decoherence can induce significant errors in the process matrix, which can then be exacerbated by the matrix logarithm. Both of these effects become increasingly problematic as the system size increases, which is why the data for the three qubit process matrices are noisier than the corresponding two qubit process matrices. This is further evidenced by the principle eigenvalue of the extracted three qubit process matrix, which is $\lambda_0=0.74$, implying a strong degree of non-unitarity. If this were due to the physical channel being non-unitary, one would expect this to also be reflected in the two qubit QPT results. However, all of the experiments showed a similar, very high degree of unitarity, with principle eigenvalues of approximately $0.93$, a value which did not significantly change with the addition of the second drive pulse. As such, it is more likely that this non-unitarity is due to measurement noise and process matrix reconstruction error rather than genuine physical processes. 

\subsection{Hamiltonian Tomography}\label{sec:ham_tomog}
Full QPT is useful for verifying that the expected dynamics are being generated in the experimental setup, but, as argued above, it is ill-suited to extracting the quantitative Hamiltonian rates necessary to characterize the system. With a verified Hamiltonian form, however, it is possible to obtain these rates far more efficiently using Hamiltonian tomography~\cite{wang2015hamiltonian,li2020hamiltonian,ma2017hamiltonian}.

For many systems of interest (including the FF transmon system which forms the basis of this work) the time evolution operator generated by the system Hamiltonian consists of a linear combination of only a small number of Pauli terms. When particularized to two qubis, the terms in the Hamiltonian Eq.~\eqref{eq:cr_ham_all_terms} form a closed group under multiplication, meaning that the cross-resonance channel can be expressed in terms of a linear combination of only the terms appearing in Eq.~\eqref{eq:cr_ham_all_terms}, rather than the full set of Pauli operators. Since the qualitative form of the Hamiltonian Eq.~\eqref{eq:cr_ham_all_terms} has been rigorously verified through QPT in the previous section, this reduced structure can be relied upon as accurate, thereby allowing the channel to be fully characterized using significantly less effort than full QPT.

We extract the rates $c_{ij}^k$ (with the exception of $c_{ij}^{Z\mathds{1}}$) in Eq.~\eqref{eq:cr_ham_all_terms} through a series of Rabi oscillation experiments acting on the $\ket{00}$ and $\ket{10}$ initial states and measuring in the $X$, $Y$ and $Z$ bases on the target qubit. The protocol for performing this set of tomography experiments may be found in the \texttt{qiskit\_experiments} framework on GitHub~\cite{qiskit_experiments}.

For pulses in which the drive envelope is purely real (i.e., $h^Y_i=0$ in Eq.~\eqref{eq:single_drive_env}), three terms are dominant: the $Z_i\mathds{1}_j$, $Z_iX_j$ and $\mathds{1}_iX_j$ terms. Similarly, for purely imaginary drive envelopes ($h^X_i=0$ in Eq.~\eqref{eq:single_drive_env}) the dynamics are dominated by $Z_i\mathds{1}_j$, $Z_iY_j$ and $\mathds{1}_iY_j$. In this case, by initializing in the $\ket{++}$ and $\ket{--}$ initial states and measuring the target qubit in the $X$ basis on the control qubit, the magnitude of the $Z_i\mathds{1}_j$ can be obtained as the sum of the frequencies of the two resulting cosine functions. With this, the full set of coefficients in Eq.~\eqref{eq:cr_ham_all_terms} may be obtained using only 8 Rabi oscillation experiments.
\begin{figure}
\centering
\subfloat[]{
\includegraphics{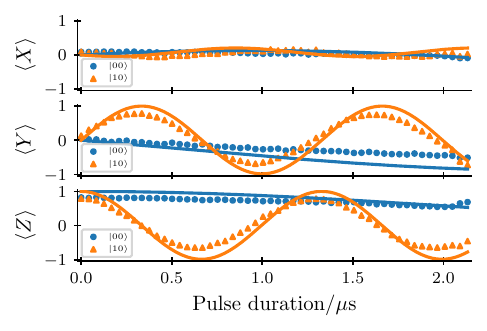}
\label{fig:ham_tomog_subfig1}}
\qquad
\subfloat[][]{
\includegraphics{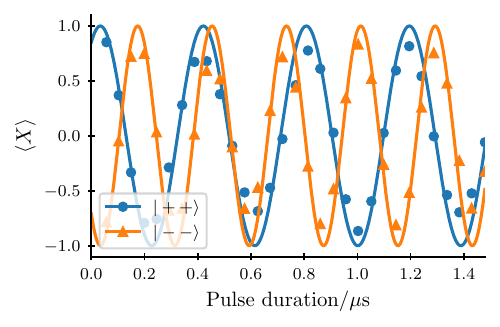}
\label{fig:ham_tomog_subfig2}}
\caption{Dynamics generated by applying the cross-resonance interaction over increasing pulse durations, with a drive amplitude of $\Omega=36$ MHz. (a) Expectation value evolution resulting from measuring qubit 2 in each Pauli basis following evolution from $\ket{00}$ (blue) and $\ket{10}$ (orange). (b) Rabi oscillations resulting from applying the cross resonance channel on the $\ket{++}$ (blue) and $\ket{--}$ (orange) initial states, measuring the first qubit. For both plots, points correspond to experimental data obtained from the \texttt{ibm\_hanoi} quantum device and the solid lines correspond to numerical expectation values extracted from evolution of the cross-resonance Hamiltonian predicted by the fits to the experimental data. The amplitudes of the oscillations are slightly reduced due to measurement error. The Hamiltonian rates obtained from these fits are given in Table.~\ref{table:hamiltonian_rates}.}
\label{fig:ham_tomog}
\end{figure}
This fitting procedure works very well, as evidenced by Fig.~\ref{fig:ham_tomog}, which shows the results of the Hamiltonian tomography implemented on the \texttt{ibm\_hanoi} quantum device. This is to be expected given the full QPT results above. The solid lines in both Fig.~\ref{fig:ham_tomog_subfig1} and Fig.~\ref{fig:ham_tomog_subfig2} are generated by numerically evolving the cross-resonance Hamiltonian obtained from the fits to the experimental data. In both cases, the points are the raw experimental data. The suppression in the observed amplitudes is most likely due to measurement error since it is present at $t=0$.
\begin{table}
\begin{center}
\begin{tabular}{|c|c|}
\hline
 Hamiltonian term & Rate/MHz \\
 \hline
 $Z_iX_j$ & -0.4915  \\
 $Z_iY_j$ & -0.0332   \\
 $Z_iZ_j$ & 0.0294   \\
 $\mathds{1}_iX_j$ & 0.4168   \\
 $\mathds{1}_iY_j$ & 0.0649   \\
 $\mathds{1}_iZ_j$ & -0.0756   \\
 $Z_i\mathds{1}_j$ & 3.0810   \\
 \hline
\end{tabular}
\caption{Cross-resonance Hamiltonian rates extracted from the \texttt{ibm\_hanoi} quantum device using Hamiltonian tomography as described in the main text, with the applied drive amplitude $\Omega=36$ MHz. The dynamics are dominated by the AC Stark shift $Z_i\mathds{1}_j$ term, which can be eliminated by adjusting the qubit drive frequency or by applying a dynamic phase to single qubit pulses, leaving dynamics generated principally by the desired $Z_iX_j$ and $\mathds{1}_iX_j$, the latter of which can be eliminated by the addition of a simultaneous resonant pulse on the target qubit. All other terms are significantly smaller. \label{table:hamiltonian_rates}}
\end{center}
\end{table}
The Hamiltonian rates obtained from the tomography are given in Table~\ref{table:hamiltonian_rates}. By far the largest term, at over $3$MHz is the $Z_i\mathds{1}_j$ term arising from the AC Stark shift on the control qubit from the off-resonant drive. This is expected to be the case based on theoretical predictions (the Stark shift term is proportional to the square of the applied drive amplitude $\Omega$, whereas the other terms are proportional to either $J_{ij}\Omega$ or $J_{ij}^2$, with $J_{ij}\ll \Omega$).

The next largest rates are the desired entangling operator $Z_iX_j$ and the spurious single qubit $\mathds{1}_iX_j$ rotation, which for these qubits and drive parameters have equal and opposite rates with magnitudes approximately $0.5$MHz. All other terms have substantially smaller magnitudes and can be neglected. In particular, the two-body $Z_iY_j$ and $Z_iZ_j$ terms, arising from residual drive phase miscalibration and qubit-qubit self-interaction respectively, are the smallest and are more than an order of magnitude smaller than the desired $Z_iX_j$ term. These are the most problematic terms, since they cannot be eliminated using single qubit quantum control.

In order to use the cross-resonance interaction for analogue quantum simulation, it is necessary to be able to control all the Hamiltonian terms. As shown in Eq.~\eqref{eq:single_drive_env}, the single qubit terms can be controlled or eliminated through the addition of resonant control pulses, while the magnitude of the two-body interactions is controlled by the amplitude of the cross-resonance drive. Since changing this amplitude also changes all the other Hamiltonian rates, a strategy for implementing analogue quantum simulation in this platform would be to fix the cross-resonance amplitude, and only use single qubit control to implement the desired simulation.

A reasonable initial target would be to generate a pure Ising-type interaction (after basis change) of the form
\begin{equation}
    H = \sum_{\langle i,j\rangle } J_{ij}^{ZX} Z_iX_j \ .
\end{equation}
In order to accomplish this, single qubit control pulses can be used to cancel the remaining spurious terms. These compensation pulses must be applied at very small amplitudes compared with those typically used for single qubit control. For comparison, the resonant amplitude for implementing a single qubit $X$ gate on an IBM Quantum device is typically $30$ MHz, approximately sixty times that needed to eliminate the spurious $\mathds{1}_iX_j$ term. It is often assumed that low amplitude drives are unproblematic, however, in real experiments implementing such small drives can cause significant problems, which will be shown in the following section.

\section{Other Sources of Error}
From the above analysis, it has been demonstrated that FF transmon qubits would be a powerful platform for analogue quantum simulation, assuming that the small pulses necessary to cancel the spurious single qubit terms can be implemented accurately. While the fidelity of single qubit dynamics is typically assumed to be much higher than that of entangling operations~\cite{smith2019simulating,vovrosh2021simple}, this breaks down in the low-amplitude regime necessitated by the analogue simulation protocol proposed here.

The weakness of the effective cross-resonance interaction also means that simulations must be run for longer times than single qubit resonant dynamics to observe interesting dynamics. As an illustrative example, consider a calibrated $Z_iX_j$ drive with a strength of $0.5$~MHz. In order to simulate the time dynamics of this system for a duration $J_{ZX}t=1$ (not an especially ambitious goal considering gate-based methods implemented on IBM Quantum devices are capable of exceeding such times for many systems of interest~\cite{vovrosh2021confinement,kapil2018quantum}) the pulses need to be applied for a duration of $2\mu$s. In principle, this is not an issue, since the reported decoherence and dephasing times for FF devices significantly exceed these times (they are typically on the order of $100 \ \mu$s~\cite{koch2020demonstrating}). Since the viability of analogue quantum simulation relies heavily upon this feature of FF transmon qubits, the reliability of these decoherence times should be verified in real experimental settings. Additionally, over such long durations, small errors can greatly reduce the fidelity of the applied dynamics, making characterization of the dynamics at low pulse amplitudes and over long times crucial for analogue quantum simulation.

\subsection{Low Amplitude Errors}\label{sec:low_amp_drive}
One problematic source of errors arising from the weak cross-resonance interaction is the deviation from theoretical predictions for low amplitude ($\Omega$ less than approximately $1.5$MHz) resonant drives. This has significant implications for the realization of the analogue simulations described here due to the weakness of the cross-resonance interaction compared to typical resonant control.

The dynamics generated by a weak resonant pulse can be investigated using single qubit quantum state tomography following the evolution of the $\ket{0}$ state over time. Given the weak nature of the pulse and the fact that any coupled qubits have frequencies which are well detuned from the drive qubit, it is reasonable to ignore spectator qubits.

Following Eq.~\eqref{eq:single_qubit_eff} and applying pulses which are approximately constant such that $h_i$ are approximately time-independent (with deviations from this primarily arising from the non-zero pulse ramp time), the resulting dynamics can be fit to a model Hamiltonian of the form
\begin{equation}\label{eq:static_single_q_ham}
    H = \frac{\Omega}{2}\left(h^XX + h^YY + h^ZZ\right) \ .
\end{equation}
For the purposes of this work, the target channel is generated by the application of a purely real pulse with zero detuning such that the effective Hamiltonian is simply $\frac{1}{2}\Omega X$. By performing the state tomography experiments over an array of pulse durations, the magnitudes of the $X, Y$ and $Z$ terms can be extracted, allowing for the phase (resulting in spurious $Y$ terms), detuning (resulting in spurious $Z$ terms) and amplitude (resulting in incorrect Rabi frequencies) errors to be characterized. Since the necessary evolution times for investigating such weak fields are long enough for dephasing and decoherence effects to be observed, an additional exponential decay of the form $\exp(-t/T_2)$ was included in the dynamics, allowing for an approximation of the $T_2$ decay time to also be extracted.
\begin{figure}
\centering
\subfloat[][]{
\includegraphics[width=0.45\textwidth]{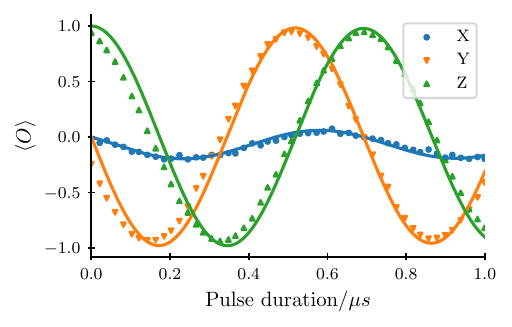}
\label{fig:low_amp_6}} \\
\subfloat[][]{
\includegraphics[width=0.45\textwidth]{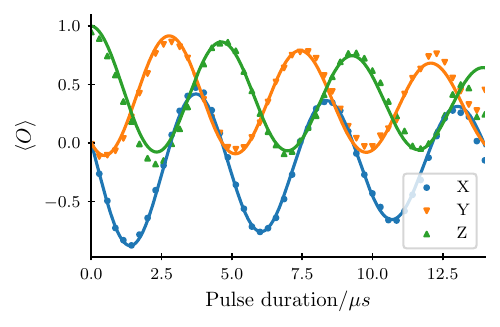}
\label{fig:low_amp_0}}
\caption{Single qubit quantum state tomography results showing $\langle X\rangle, \langle Y\rangle$ and $\langle Z\rangle$ expectation values over varying drive durations, with drive amplitudes of (a) 1.50 MHz and (b) 0.07 MHz, implemented on the \texttt{ibm\_hanoi} quantum device. The points correspond to experimental data and the solid lines correspond to fits to Eq.~\eqref{eq:static_single_q_ham}, from which phase, detuning and amplitude errors may be extracted. For (a), the observed Rabi frequency was $1.44$MHz, the observed phase error was $-0.04\pi$ and the observed detuning was $0.05$MHz. For (b), the observed Rabi frequency was $0.16$MHz, the observed phase error was $0.65\pi$ and the observed detuning was $0.07$MHz. The observed $T_2$ time for both (a) and (b) was $31.69\mu$s (this was extracted from (b) and used in the plots of both (a) and (b)).}
\label{fig:low_amplitude_errors}
\end{figure}
Fig.~\ref{fig:low_amplitude_errors} shows the results of such a state tomography scheme for two low amplitudes, 0.07 MHz, corresponding to $0.0005\%$ of the maximum drive amplitude accessible through the IBM Pulse platform and $1.50$MHz, corresponding to $0.01\%$ of the maximum drive amplitude. It should be stressed that although these amplitudes are very small when compared with typical resonant control pulses on FF qubits, it is often necessary to use such low amplitudes for cancelling the spurious terms in the cross-resonance Hamiltonian.

For both Fig.~\ref{fig:low_amp_0}~and~Fig.~\ref{fig:low_amp_6}, there are significant amplitude, phase and detuning errors, with the relative size of the errors being much higher for the lower drive amplitude. For Fig.~\ref{fig:low_amp_6}, the experimental Rabi frequency (obtained as $\sqrt{(h^X)^2+(h^Y)^2}$) was observed to be $1.44$MHz, corresponding to an error of $0.06$MHz or 4\%. The phase error (obtained by fitting $h^X$ and $h^Y$ to Fig.~\ref{fig:low_amp_6} and then obtaining the phase of $h^X + ih^Y$) was found to be $-0.04$ with the detuning error being $0.05$MHz. The origin of these errors is most likely due to miscalibration: since the typical target amplitudes for implementing quantum gates are as high as possible without inducing unwanted transitions, the experimental settings for the resonant drives will be calibrated for this regime. The low amplitudes used in these experiments are far from this regime, and so moderate non-linearities in the control hardware could result in deviations between the software and the experimental realization. While these deviations are not ideal, and could cause problems for an experimental simulation in which precise parameters are required, in principle the errors can be counteracted by adjusting the amplitude, phase and frequency of the applied pulse.

For Fig.~\ref{fig:low_amp_0}, the experimental errors are much higher, with the observed Rabi frequency being $0.16$~MHz, corresponding to an error of $0.09$~MHz or $129\%$ of the theoretical drive. The phase error was also significantly higher, at approximately $0.65\pi$, while the observed detuning was $0.07$~MHz. The latter two errors, although higher than Fig.~\ref{fig:low_amp_6}, should also be able to be corrected in the same way. The amplitude error, however, cannot be fixed in software, as it likely arises from the finite resolution of the arbitrary waveform generators (AWGs). AWGs with more than the required resolution can be built using commercially-available components~\cite{lin2019scalable}. Thus, while this is a limiting factor preventing the implementation of analogue quantum simulations on currently available devices, it should not represent an insurmountable challenge for such an implementation in the near future.

\subsection{Spurious Dynamics in the Undriven System}\label{sec:tls_coupling}
A second consequence of the weak cross-resonance interaction is that the simulations need to be performed for longer times than for resonant single qubit dynamics (on the order of approximately $10 \mu s$). Both the reported $T_2$ and bare $T_2^*$ dephasing times for the IBM Quantum devices should be long enough for this to be achieved~\cite{koch2020demonstrating}. However, experimentally, additional spurious terms may be observed which significantly impact the behaviour of FF transmon systems over this time period. This is illustrated when one attempts to measure the dephasing of the system experimentally.
\begin{figure}
    \includegraphics{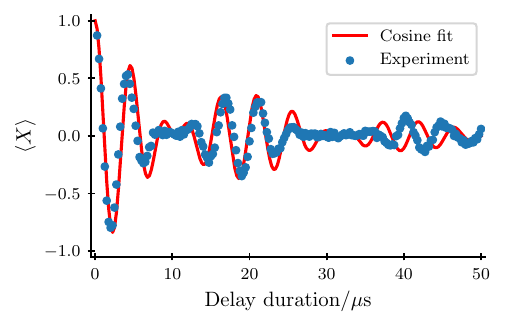}
    \caption{Plot of spurious dynamics observed on the \texttt{ibm\_hanoi} quantum device in the absence of driving as a function of delay time, with the points corresponding to the experimental data and the red line corresponding to a least squares best fit to Eq.~\eqref{eq:cosine_fit}. The qubit is initialised in the $\ket{+}$ $X$ eigenstate, left idle for varying durations and measured in the $X$ basis. The same dynamics were not observed for the same experiment performed in the $Z$ basis, indicating the presence of a spurious $Z$ field. Given the structure of the observed interaction, the most likely explanation is that it is caused by coupling to a two level system defect in the superconducting material.}
    \label{fig:dephasing_test_2}
\end{figure}
The dephasing of a quantum system can be experimentally measured by initializing in the $\ket{+}$ state, waiting for varying durations, and then measuring in the $X$ eigenbasis. In a system with dephasing error, the expectation value should exhibit an exponential decay over the characteristic $T_2^*$ time. In the absence of any driving or spurious fields, there should not be any oscillatory behaviour. This is not necessarily the case experimentally. Fig.~\ref{fig:dephasing_test_2} shows the results of performing such an experiment on the \texttt{ibm\_hanoi} quantum device, with the results being dominated by strong oscillations. This effect is not symmetric for all axes: performing the same experiment in the $Z$ eigenbasis results in no such oscillations, implying that the effect arises due to a spurious $Z$ field.

The oscillations in Fig.~\ref{fig:dephasing_test_2} may be fit to a function of the form
\begin{equation}\label{eq:cosine_fit}
    f(t) = \left(c_0\cos(f_0 t) + c_1\cos(f_1 t)\right)e^{-t/T_2^*} \ ,
\end{equation}
with five fit parameters $\{c_0, c_1, f_0, f_1, T_2^*\}$. The results of such a fit for qubit 16 in the \texttt{ibm\_hanoi} quantum device are given as the red line in Fig.~\ref{fig:dephasing_test_2}. For this fit, the optimal frequencies were obtained as $f_0=0.18$ MHz and $f_1=0.24$ MHz, with weight coeffients $c_0=0.48$ and $c_1=0.51$. These coefficients do not reproduce the behaviour observed when measuring other qubits or when measuring the same qubit on different days and thus should be considered to be specific to this qubit at this particular time. As expected, the $T_2^*$ time was found to be significantly shorter than the $T_2$ time reported by IBM, at $20.2\mu$s as opposed to $209.6\mu$s for this qubit. This may still be long enough to offer some advantage for analogue quantum simulation.

Since there are two observed frequencies in the oscillations, the observed dynamics cannot be explained by a simple frequency misalignment. While it is difficult to identify the source of these errors with any certainty given the limited experimental access available to end users, the form of the dynamics is consistent with coupling between the qubit and a mesoscopic environment. Such an environment could be provided by the presence of parasitic two-level systems (TLSs)~\cite{de2020two,lisenfeld2015observation} arising from structural defects in the superconducting material. In the experiment presented in Fig.~\ref{fig:dephasing_test_2}, coupling to a single TLS defect would induce the observed dynamics. The presence of TLSs in superconducting qubits is well known to be a significant source of decoherence~\cite{muller2011simulating}, and could be generated through background radioactivity or cosmic rays~\cite{mcewen2022resolving}. It is likely that the TLS coupling also induces significant decoherence, reducing the observed $T_2^*$ time. Eliminating this coupling could therefore significantly increase the decoherence times, allowing for longer analogue simulation times to be reached and extending the utility of such protocols.
\begin{figure}
    \includegraphics{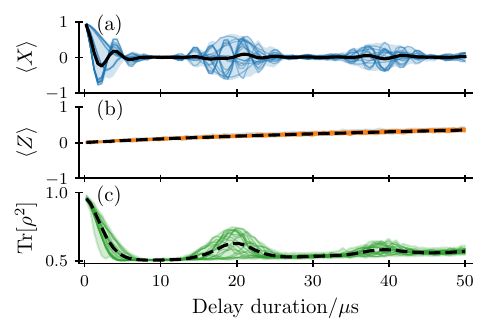}
    \caption{Single qubit time dynamics and purity measurements corresponding to experiments initialized in $|+\rangle$ and measured in all three Pauli bases following different delay times, with the experiments implemented on the \texttt{ibm\_hanoi} quantum device. The data were obtained from the same qubit as Fig.~\ref{fig:dephasing_test_2}, but on a different day. (a) Time evolution of $\langle X\rangle$ expectation values, with the black line corresponding to the mean evolution, the blue lines corresponding to experimental evolution and the shaded region bounding the highest and lowest points observed. (b) Time evolution of $\langle Z\rangle$ expectation values, with the black line corresponding to the mean evolution, the orange lines corresponding to experimental evolution and the shaded region bounding the highest and lowest points observed. (c) Purity measurements for the delay experiments. While a multi-frequency classical field together with $T_2$ dephasing could explain the results of (a) and (b), the revivals in the purity at approximately $20\mu$s and $40\mu$s in (c) cannot be explained by classical noise sources.}
    \label{fig:xyz_purity_delay}
\end{figure}
The nature of the spurious dynamics varies significantly over time. Fig.~\ref{fig:xyz_purity_delay}(a) shows 100 repetitions of the dephasing test outlined above, obtained on a different day to the data presented in Fig.~\ref{fig:dephasing_test_2} but for the same qubit and device. The blue lines show individual evolution for a characteristic subset of 20 experiments, with the blue shaded region corresponding to the upper and lower bounds of the obtained results. These data also show similar oscillatory behaviour, but cannot be well reproduced by fits to Eq.~\eqref{eq:cosine_fit}. This can be explained by the qubit coupling to multiple TLS defects rather than a single one as in Fig.~\ref{fig:dephasing_test_2}. Moreover, the oscillations fluctuate significantly over the time scale during which the experiments were performed (approximately six hours), indicating that the cause of the oscillatory behaviour changes substantially on this time scale. Performing the same experiment, but measuring in the $Z$ basis (Fig.~\ref{fig:xyz_purity_delay}(b)) yields no oscillations for any of the 100 experimental repetitions, providing further evidence that the dynamics are induced by an effective $Z$ field. The increase in the $\langle Z \rangle$ expectation value over time arises from the decay of the initial state $|+\rangle$ to a linear combination of $|+\rangle$ and $|-\rangle$ over time due to dephasing.

The black line in Fig.~\ref{fig:xyz_purity_delay}(a) shows the average behaviour across all experimental runs, with the limiting behaviour trending towards an exponential decay. Performing the delay experiment on different qubits on different days does not always result in oscillations, but can also result in decay behaviour similar to this mean behaviour. If the variation in the oscillatory dynamics can be explained by coupling to a mesoscopic TLS environment in which the number of defects is not constant over time, then the time-average behaviour of this system is likely to be qualitatively similar to coupling between a qubit and a bulk environment of defects. This decay behaviour could therefore be plausibly explained by coupling to a macroscopic environment of TLS defects.

The identification of the source of the oscillations with coupling to a low-dimensional environment is further reinforced by Fig.~\ref{fig:xyz_purity_delay}(c), which shows the purity of the single qubit state as a function of delay time for the same experiments as Figs.~\ref{fig:xyz_purity_delay}(a)~and~(b). The purity measurements show a significant revival at approximately $20 \mu$s and a smaller revival at approximately $40 \mu$s. Such revivals cannot be explained by classical sources of noise (such as, for example, magnetic fields induced by nearby power lines) and are indicative of coherent coupling between the qubit and some other quantum object. This is further evidence in support of the identification of the spurious dynamics with coupling to a low-dimensional environment consisting of TLS defects.

There is therefore strong evidence that the observed dynamics are due to a coherent coupling process between the qubit and some other quantum object. Based on the form of the oscillations and the fact that TLSs have been identified as a major source of decoherence in other superconducting platforms, it is reasonable to identify this quantum object with TLS defects. However, this cannot be rigorously verified (nor, indeed, disproven) using the level of experimental access available to end users. The spurious oscillating terms, regardless of their origin, are a significant barrier to the implementation of analogue quantum simulation on FF transmon qubits and thus the experimental investigation of their origin is of crucial importance if such a goal is to be achieved.

\section{Assessment against the viability criteria}
In this section, the conclusions drawn from the experimental results presented in the rest of the paper are used to explicitly evaluate the performance of FF transmon qubits with respect to the criteria of (C1) expressibility, (C2) practical controllability and (C3) stability outlined in the introduction.

\subsection*{(C1) Expressibility}
As shown in Sec.~\ref{sec:background}, FF transmons are, theoretically, highly expressible, and should therefore satisfy this criterion. Through the full quantum process tomography results in Sec.~\ref{sec:qpt}, the validity of Eq.~\eqref{eq:cr_ham_all_terms} (and therefore the expressibility of the system) is verified. Thus, criterion (C1) is satisfied.

\subsection*{(C2) Practical Controllability}
The full process tomography results for different driving setups in Sec.~\ref{sec:qpt} show that the two-qubit coupling terms can be applied in parallel without inducing significant additional noise. Additionally, the Hamiltonian tomography results in Sec.~\ref{sec:ham_tomog} show that a calibration procedure to obtain the amplitudes necessary to control all the Hamiltonian terms is feasible, being able to be completed with greatly reduced effort compared with full quantum process tomography. These experiments imply that the FF transmon system is indeed controllable -- however, the issues with low amplitude single qubit driving presented in Sec.~\ref{sec:low_amp_drive} mean that full control is not possible on currently available devices. This is mostly attributed to the finite resolution of the control hardware not allowing for the accurate implementation of low amplitude pulses, which are themselves necessary due to the relatively weak cross-resonance interaction compared with typical resonant single-qubit drive strengths. Such a limitation could be overcome with moderate hardware improvements. Alternatively, analogue control schemes could be devised that incorporate the control limitations into the simulation (for example, one could simulate Hamiltonian dynamics in the presence of a static magnetic field). Thus, criterion (C2) is not currently satisfied on available devices, but it is close to being satisfied and the limitations imposed by the lack of practical controllability could be overcome with moderate effort.

\subsection*{(C3) Stability}
One of the most favorable aspects of FF transmon qubits as a platform for quantum computing is their long coherence times. In Sec.~\ref{sec:ham_tomog}~and~\ref{sec:low_amp_drive}, these long coherence times are observed for both the two-qubit Hamiltonian tomography and the single qubit driving results respectively, providing experimental verification of this property. However, the stability of the experimental system is called into question by the evidence of TLS coupling presented in Sec.~\ref{sec:tls_coupling}. The presence of this coupling could, in principle, be accounted for in the control scheme for an analogue simulation. Although approaches for achieving such control over the defects are still in their infancy, with robust experimental techniques for characterising TLS defects in superconducting materials only being published in the last few years~\cite{lisenfeld2015observation,de2020two,abdurakhimov2022identification}, practical approaches have been proposed using techniques such as noise spectrum engineering~\cite{you2022qcb}. The more problematic feature of this coupling is its transient nature, changing dramatically over relatively short periods of time, which could cause issues for analogue simulations that could not be alleviated through a calibration procedure. For this reason, criterion (C3) is not currently satisfied.

\section{Conclusion and Outlook}
Quantum simulation is one of the flagship applications of quantum devices. Current digital devices satisfy almost all of the criteria necessary for the realization of such simulations, but gate errors continue to pose significant problems for digital simulation methods such as Trotterization.

Analogue quantum simulation is an alternative route towards realizing Hamiltonian simulation. We experimentally evaluate fixed-frequency transmon qubits against three criteria which must be satisfied in order to realise analogue quantum simulation experimentally: (C1) expressibility, (C2) practical controllability and (C3) stability.

Our results indicate that superconducting fixed-frequency devices are a flexible and highly controllable analogue simulation platform, and we demonstrate that the Hamiltonian of the system can be measured quickly and with high confidence.

We find that the weakness of the cross-resonance interaction causes significant issues. At the low amplitudes required to counteract the spurious single qubit terms arising from the cross-resonance interaction, significant phase, amplitude and detuning errors cause problematic deviations from the ideal resonant drive. At moderate drive amplitudes these can be addressed on currently available devices by calibrating the drive pulses, but for very low amplitudes the limited resolution of the arbitrary waveform generators (AWGs) precludes such calibration protocols from being realized. This issue could be easily addressed using high resolution AWGs, and thus is not a significant limiting factor for the implementation of analogue quantum simulations on FF transmon qubits. Criteria (C1) (the system Hamiltonian must be expressible enough to permit simulation of systems of interest) and (C2) (the system must be controllable such that individual Hamiltonian terms can be switched on and off independently) may then be satisfied with only modest improvements to current FF transmon implementations. 

Alternatively, the floor on the implementable drive pulse amplitudes could be interpreted as a minimum disorder term for analogue simulations. For the example drives here, the minimum disorder would correspond to approximately one third of the $Z_iX_j$ interaction strength, although the precise value varies from qubit to qubit.

A more problematic issue is diagnosed through the presence of dynamical evolution on qubits which are initialized in an $X$ eigenstate and left without additional drive pulse. We attribute the observed dynamics to coupling to two level system defects, which is known to be a major source of decoherence in other superconducting platforms. Since the spurious dynamics fluctuate over time, any protocol designed to eliminate the interaction would need to account for this fluctuation in order to ensure that criterion (C3) (the system must remain stable over the duration of the simulation) is satisfied.

Since rigorous verification and full characterization of the TLS interaction are not possible with the level of experimental access available to end users, a crucial route for further investigation lies in the application of spectroscopic techniques reported in other superconducting quantum platforms for the full characterization of this effect~\cite{dong2022measurement} and in the development of an effective method to eliminate these interactions. While the focus of this work is on evaluating FF qubits as a platform for analogue quantum simulation, this particular error could have a significant impact on gate-based computation as well. TLS coupling has been identified as a key source of decoherence in superconducting platforms~\cite{muller2011simulating}, and so a focus on addressing this problem could have wide-reaching implications beyond analogue simulation.

In summary, FF transmon qubits have great potential as a platform for analogue quantum simulation due to their ability to implement a wide array of physically interesting Hamiltonians. The primary limiting factors preventing the implementation of such simulations on current devices are the limited resolution of the control pulses and spurious coupling to two level system defects. The former can be straightforwardly eliminated with either modest hardware improvements or incorporation of the limitations into the simulations, while the latter requires further experimental work identify, characterize and eliminate the coupling. Overcoming these limitations will allow for the realization of the potential of FF transmon qubits for analogue quantum simulation.

\section{Acknowledgments}
We are grateful to Johannes Knolle, Kiran Khosla and Francesco Petiziol for providing stimulating discussions. This work is supported by Samsung GRP grant, the UK Hub in Quantum Computing and Simulation, part of the UK National Quantum Technologies Programme with funding from UKRI EPSRC grant EP/T001062/1 and the QuantERA ERA-NET Co-fund in Quantum Technologies implemented within the European Union’s Horizon 2020 Programme. This research is part of the Munich Quantum Valley, which is supported by the Bavarian state government with funds from the Hightech Agenda Bayern Plus. S.G. is supported by a studentship in the Quantum Systems Engineering Skills and Training Hub at Imperial College London funded by EPSRC (EP/P510257/1). D.M. acknowledges funding from ERC Advanced Grant QUENOCOBA under the EU Horizon 2020 program (Grant Agreement No. 742102). A.S. was supported by a research fellowship from the The Royal Commission for the Exhibition of 1851. We acknowledge the use of IBM Quantum services for this work. The views expressed are those of the authors, and do not reflect the official policy or position of IBM or the IBM Quantum team.

\bibliography{bibliography.bib}

\begin{thebibliography}{48}
\providecommand{\natexlab}[1]{#1}
\providecommand{\url}[1]{\texttt{#1}}
\expandafter\ifx\csname urlstyle\endcsname\relax
  \providecommand{\doi}[1]{doi: #1}\else
  \providecommand{\doi}{doi: \begingroup \urlstyle{rm}\Url}\fi

\bibitem[Abdurakhimov et~al.(2022)Abdurakhimov, Mahboob, Toida, Kakuyanagi, Matsuzaki, and Saito]{abdurakhimov2022identification}
Leonid~V. Abdurakhimov, Imran Mahboob, Hiraku Toida, Kosuke Kakuyanagi, Yuichiro Matsuzaki, and Shiro Saito.
\newblock Identification of different types of high-frequency defects in superconducting qubits.
\newblock \emph{PRX Quantum}, 3:\penalty0 040332, Dec 2022.
\newblock \doi{10.1103/PRXQuantum.3.040332}.
\newblock URL \url{10.1103/PRXQuantum.3.040332}.

\bibitem[ANIS et~al.()ANIS, Abby-Mitchell, Abraham, AduOffei, Agarwal, Agliardi, Aharoni, Ajith, Akhalwaya, Aleksandrowicz, et~al.]{qiskit_experiments}
MD~SAJID ANIS, Abby-Mitchell, H{\'e}ctor Abraham, AduOffei, Rochisha Agarwal, Gabriele Agliardi, Merav Aharoni, Vishnu Ajith, Ismail~Yunus Akhalwaya, Gadi Aleksandrowicz, et~al.
\newblock Qiskit experiments, available at github.com/qiskit/qiskit-experiments.
\newblock URL \url{https://github.com/Qiskit/qiskit-experiments.git}.

\bibitem[ANIS et~al.(2021)ANIS, Abby-Mitchell, Abraham, AduOffei, Agarwal, Agliardi, Aharoni, Ajith, Akhalwaya, Aleksandrowicz, et~al.]{Qiskit2}
MD~SAJID ANIS, Abby-Mitchell, H{\'e}ctor Abraham, AduOffei, Rochisha Agarwal, Gabriele Agliardi, Merav Aharoni, Vishnu Ajith, Ismail~Yunus Akhalwaya, Gadi Aleksandrowicz, et~al.
\newblock Qiskit: An open-source framework for quantum computing, 2021.

\bibitem[Arute et~al.(2019)Arute, Arya, Babbush, Bacon, Bardin, Barends, Biswas, Boixo, Brandao, Buell, et~al.]{arute2019quantum}
Frank Arute, Kunal Arya, Ryan Babbush, Dave Bacon, Joseph~C Bardin, Rami Barends, Rupak Biswas, Sergio Boixo, Fernando~GSL Brandao, David~A Buell, et~al.
\newblock Quantum supremacy using a programmable superconducting processor.
\newblock \emph{Nature}, 574\penalty0 (7779):\penalty0 505--510, 2019.
\newblock \doi{10.1038/s41586-019-1666-5}.

\bibitem[Barends et~al.(2016)Barends, Shabani, Lamata, Kelly, Mezzacapo, Heras, Babbush, Fowler, Campbell, Chen, et~al.]{barends2016digitized}
Rami Barends, Alireza Shabani, Lucas Lamata, Julian Kelly, Antonio Mezzacapo, U~Las Heras, Ryan Babbush, Austin~G Fowler, Brooks Campbell, Yu~Chen, et~al.
\newblock Digitized adiabatic quantum computing with a superconducting circuit.
\newblock \emph{Nature}, 534\penalty0 (7606):\penalty0 222--226, 2016.
\newblock \doi{10.1038/nature17658}.

\bibitem[Blais et~al.(2020)Blais, Girvin, and Oliver]{blais2020quantum}
Alexandre Blais, Steven~M Girvin, and William~D Oliver.
\newblock Quantum information processing and quantum optics with circuit quantum electrodynamics.
\newblock \emph{Nat. Phys.}, 16\penalty0 (3):\penalty0 247--256, 2020.
\newblock \doi{10.1038/s41567-020-0806-z}.

\bibitem[Blatt and Roos(2012)]{blatt2012quantum}
Rainer Blatt and Christian~F Roos.
\newblock Quantum simulations with trapped ions.
\newblock \emph{Nat. Phys.}, 8\penalty0 (4):\penalty0 277--284, 2012.
\newblock \doi{10.1038/nphys2252}.

\bibitem[Browaeys and Lahaye(2020)]{browaeys2020many}
Antoine Browaeys and Thierry Lahaye.
\newblock Many-body physics with individually controlled {R}ydberg atoms.
\newblock \emph{Nat. Phys.}, 16\penalty0 (2):\penalty0 132--142, 2020.
\newblock \doi{10.1038/s41567-019-0733-z}.

\bibitem[Chow et~al.(2011)Chow, C{\'o}rcoles, Gambetta, Rigetti, Johnson, Smolin, Rozen, Keefe, Rothwell, Ketchen, et~al.]{chow2011simple}
Jerry~M Chow, Antonio~D C{\'o}rcoles, Jay~M Gambetta, Chad Rigetti, Blake~R Johnson, John~A Smolin, Jim~R Rozen, George~A Keefe, Mary~B Rothwell, Mark~B Ketchen, et~al.
\newblock Simple all-microwave entangling gate for fixed-frequency superconducting qubits.
\newblock \emph{Phys. Rev. Lett.}, 107\penalty0 (8):\penalty0 080502, 2011.
\newblock \doi{10.1103/PhysRevLett.107.080502}.

\bibitem[Cirac and Zoller(2012)]{cirac2012goals}
J~Ignacio Cirac and Peter Zoller.
\newblock Goals and opportunities in quantum simulation.
\newblock \emph{Nat. Phys.}, 8\penalty0 (4):\penalty0 264--266, 2012.
\newblock \doi{10.1038/nphys2275}.

\bibitem[de~Graaf et~al.(2020)de~Graaf, Faoro, Ioffe, Mahashabde, Burnett, Lindstr{\"o}m, Kubatkin, Danilov, and Tzalenchuk]{de2020two}
SE~de~Graaf, L~Faoro, LB~Ioffe, S~Mahashabde, JJ~Burnett, T~Lindstr{\"o}m, SE~Kubatkin, AV~Danilov, and A~Ya Tzalenchuk.
\newblock Two-level systems in superconducting quantum devices due to trapped quasiparticles.
\newblock \emph{Sci. Adv.}, 6\penalty0 (51):\penalty0 eabc5055, 2020.
\newblock \doi{10.1126/sciadv.abc5055}.

\bibitem[DiVincenzo(2000)]{divincenzo2000physical}
David~P DiVincenzo.
\newblock The physical implementation of quantum computation.
\newblock \emph{Fortschr. Phys.}, 48\penalty0 (9-11):\penalty0 771--783, 2000.
\newblock \doi{10.1002/1521-3978(200009)48:9/11<771::AID-PROP771>3.0.CO;2-E}.

\bibitem[Dong et~al.(2022)Dong, Li, Zheng, Zhang, Ma, Tan, and Yu]{dong2022measurement}
Yuqian Dong, Yong Li, Wen Zheng, Yu~Zhang, Zhuang Ma, Xinsheng Tan, and Yang Yu.
\newblock Measurement of quasiparticle diffusion in a superconducting transmon qubit.
\newblock \emph{Appl. Sci.}, 12\penalty0 (17):\penalty0 8461, 2022.
\newblock \doi{10.3390/app12178461}.

\bibitem[Endres et~al.(2011)Endres, Cheneau, Fukuhara, Weitenberg, Schauss, Gross, Mazza, Banuls, Pollet, Bloch, et~al.]{endres2011observation}
Manuel Endres, Marc Cheneau, Takeshi Fukuhara, Christof Weitenberg, Peter Schauss, Christian Gross, Leonardo Mazza, Mari~Carmen Banuls, L~Pollet, Immanuel Bloch, et~al.
\newblock Observation of correlated particle-hole pairs and string order in low-dimensional {M}ott insulators.
\newblock \emph{Science}, 334\penalty0 (6053):\penalty0 200--203, 2011.
\newblock \doi{10.1126/science.1209284}.

\bibitem[Georgescu et~al.(2014)Georgescu, Ashhab, and Nori]{georgescu2014quantum}
Iulia~M Georgescu, Sahel Ashhab, and Franco Nori.
\newblock Quantum simulation.
\newblock \emph{Rev. Mod. Phys.}, 86\penalty0 (1):\penalty0 153, 2014.
\newblock \doi{10.1103/RevModPhys.86.153}.

\bibitem[Greif et~al.(2013)Greif, Uehlinger, Jotzu, Tarruell, and Esslinger]{greif2013short}
Daniel Greif, Thomas Uehlinger, Gregor Jotzu, Leticia Tarruell, and Tilman Esslinger.
\newblock Short-range quantum magnetism of ultracold fermions in an optical lattice.
\newblock \emph{Science}, 340\penalty0 (6138):\penalty0 1307--1310, 2013.
\newblock \doi{10.1126/science.1236362}.

\bibitem[Greiner et~al.(2002)Greiner, Mandel, Esslinger, H{\"a}nsch, and Bloch]{greiner2002quantum}
Markus Greiner, Olaf Mandel, Tilman Esslinger, Theodor~W H{\"a}nsch, and Immanuel Bloch.
\newblock Quantum phase transition from a superfluid to a {M}ott insulator in a gas of ultracold atoms.
\newblock \emph{Nature}, 415\penalty0 (6867):\penalty0 39--44, 2002.
\newblock \doi{10.1038/415039a}.

\bibitem[Hartmann(2016)]{hartmann2016quantum}
Michael~J Hartmann.
\newblock Quantum simulation with interacting photons.
\newblock \emph{J. Opt.}, 18\penalty0 (10):\penalty0 104005, 2016.
\newblock \doi{10.1088/2040-8978/18/10/104005}.

\bibitem[Hartmann et~al.(2008)Hartmann, Brandao, and Plenio]{hartmann2008quantum}
Michael~J Hartmann, Fernando~GSL Brandao, and Martin~B Plenio.
\newblock Quantum many-body phenomena in coupled cavity arrays.
\newblock \emph{Laser Photonics Rev.}, 2\penalty0 (6):\penalty0 527--556, 2008.
\newblock \doi{10.1002/lpor.200810046}.

\bibitem[Houck et~al.(2012)Houck, T{\"u}reci, and Koch]{houck2012chip}
Andrew~A Houck, Hakan~E T{\"u}reci, and Jens Koch.
\newblock On-chip quantum simulation with superconducting circuits.
\newblock \emph{Nat. Phys.}, 8\penalty0 (4):\penalty0 292--299, 2012.
\newblock \doi{10.1038/nphys2251}.

\bibitem[Kapil et~al.(2018)Kapil, Behera, and Panigrahi]{kapil2018quantum}
Manik Kapil, Bikash~K Behera, and Prasanta~K Panigrahi.
\newblock Quantum simulation of {K}lein {G}ordon equation and observation of klein paradox in {IBM} quantum computer.
\newblock \emph{arXiv preprint arXiv:1807.00521}, 2018.
\newblock \doi{10.48550/arXiv.1807.00521}.

\bibitem[Koch et~al.(2020)Koch, Martin, Patel, Wessing, and Alsing]{koch2020demonstrating}
Daniel Koch, Brett Martin, Saahil Patel, Laura Wessing, and Paul~M Alsing.
\newblock Demonstrating {NISQ} era challenges in algorithm design on {IBM}'s 20 qubit quantum computer.
\newblock \emph{AIP Adv.}, 10\penalty0 (9):\penalty0 095101, 2020.
\newblock \doi{10.1063/5.0015526}.

\bibitem[Krantz et~al.(2019)Krantz, Kjaergaard, Yan, Orlando, Gustavsson, and Oliver]{krantz2019quantum}
Philip Krantz, Morten Kjaergaard, Fei Yan, Terry~P Orlando, Simon Gustavsson, and William~D Oliver.
\newblock A quantum engineer's guide to superconducting qubits.
\newblock \emph{Appl. Phys. Rev.}, 6\penalty0 (2):\penalty0 021318, 2019.
\newblock \doi{10.1063/1.5089550}.

\bibitem[Lanyon et~al.(2011)Lanyon, Hempel, Nigg, M{\"u}ller, Gerritsma, Z{\"a}hringer, Schindler, Barreiro, Rambach, Kirchmair, et~al.]{lanyon2011universal}
Ben~P Lanyon, Cornelius Hempel, Daniel Nigg, Markus M{\"u}ller, Rene Gerritsma, F~Z{\"a}hringer, Philipp Schindler, Julio~T Barreiro, Markus Rambach, Gerhard Kirchmair, et~al.
\newblock Universal digital quantum simulation with trapped ions.
\newblock \emph{Science}, 334\penalty0 (6052):\penalty0 57--61, 2011.
\newblock \doi{10.1126/science.1208001}.

\bibitem[Li et~al.(2020)Li, Zou, and Hsieh]{li2020hamiltonian}
Zhi Li, Liujun Zou, and Timothy~H Hsieh.
\newblock Hamiltonian tomography via quantum quench.
\newblock \emph{Phys. Rev. Lett.}, 124\penalty0 (16):\penalty0 160502, 2020.
\newblock \doi{10.1103/PhysRevLett.124.160502}.

\bibitem[Lin et~al.(2019)Lin, Liang, Xu, Sun, Guo, Liao, and Peng]{lin2019scalable}
Jin Lin, Fu-Tian Liang, Yu~Xu, Li-Hua Sun, Cheng Guo, Sheng-Kai Liao, and Cheng-Zhi Peng.
\newblock Scalable and customizable arbitrary waveform generator for superconducting quantum computing.
\newblock \emph{AIP Adv.}, 9\penalty0 (11):\penalty0 115309, 2019.
\newblock \doi{10.1063/1.5120299}.

\bibitem[Lisenfeld et~al.(2015)Lisenfeld, Grabovskij, M{\"u}ller, Cole, Weiss, and Ustinov]{lisenfeld2015observation}
J{\"u}rgen Lisenfeld, Grigorij~J Grabovskij, Clemens M{\"u}ller, Jared~H Cole, Georg Weiss, and Alexey~V Ustinov.
\newblock Observation of directly interacting coherent two-level systems in an amorphous material.
\newblock \emph{Nat. Commun.}, 6\penalty0 (1):\penalty0 1--6, 2015.
\newblock \doi{10.1038/ncomms7182}.

\bibitem[Lloyd(1996)]{lloyd1996universal}
Seth Lloyd.
\newblock Universal quantum simulators.
\newblock \emph{Science}, 273\penalty0 (5278):\penalty0 1073--1078, 1996.
\newblock \doi{10.1126/science.273.5278.1073}.

\bibitem[Ma et~al.(2017)Ma, Owens, LaChapelle, Schuster, and Simon]{ma2017hamiltonian}
Ruichao Ma, Clai Owens, Aman LaChapelle, David~I Schuster, and Jonathan Simon.
\newblock Hamiltonian tomography of photonic lattices.
\newblock \emph{Phys. Rev. A}, 95\penalty0 (6):\penalty0 062120, 2017.
\newblock \doi{10.1103/PhysRevA.95.062120}.

\bibitem[Malekakhlagh et~al.(2020)Malekakhlagh, Magesan, and McKay]{malekakhlagh2020first}
Moein Malekakhlagh, Easwar Magesan, and David~C McKay.
\newblock First-principles analysis of cross-resonance gate operation.
\newblock \emph{Phys. Rev. A}, 102\penalty0 (4):\penalty0 042605, 2020.
\newblock \doi{10.1103/PhysRevA.102.042605}.

\bibitem[Malz and Smith(2021)]{malz2021topological}
Daniel Malz and Adam Smith.
\newblock Topological two-dimensional {F}loquet lattice on a single superconducting qubit.
\newblock \emph{Phys. Rev. Lett.}, 126\penalty0 (16):\penalty0 163602, 2021.
\newblock \doi{10.1103/PhysRevLett.126.163602}.

\bibitem[McEwen et~al.(2022)McEwen, Faoro, Arya, Dunsworth, Huang, Kim, Burkett, Fowler, Arute, Bardin, et~al.]{mcewen2022resolving}
Matt McEwen, Lara Faoro, Kunal Arya, Andrew Dunsworth, Trent Huang, Seon Kim, Brian Burkett, Austin Fowler, Frank Arute, Joseph~C Bardin, et~al.
\newblock Resolving catastrophic error bursts from cosmic rays in large arrays of superconducting qubits.
\newblock \emph{Nat. Phys.}, 18\penalty0 (1):\penalty0 107--111, 2022.
\newblock \doi{10.1038/s41567-021-01432-8}.

\bibitem[M{\"u}ller et~al.(2011)M{\"u}ller, Hammerer, Zhou, Roos, and Zoller]{muller2011simulating}
M~M{\"u}ller, Klemens Hammerer, YL~Zhou, Christian~F Roos, and P~Zoller.
\newblock Simulating open quantum systems: {F}rom many-body interactions to stabilizer pumping.
\newblock \emph{New Journal of Physics}, 13\penalty0 (8):\penalty0 085007, 2011.
\newblock \doi{10.1088/1367-2630/13/8/085007}.

\bibitem[Pancotti et~al.(2020)Pancotti, Giudice, Cirac, Garrahan, and Banuls]{pancotti2020quantum}
Nicola Pancotti, Giacomo Giudice, J~Ignacio Cirac, Juan~P Garrahan, and Mari~Carmen Banuls.
\newblock Quantum {E}ast model: {L}ocalization, nonthermal eigenstates, and slow dynamics.
\newblock \emph{Phys. Rev. X}, 10\penalty0 (2):\penalty0 021051, 2020.
\newblock \doi{10.1103/PhysRevX.10.021051}.

\bibitem[Peng et~al.(2005)Peng, Du, and Suter]{peng2005quantum}
Xinhua Peng, Jiangfeng Du, and Dieter Suter.
\newblock Quantum phase transition of ground-state entanglement in a heisenberg spin chain simulated in an {NMR} quantum computer.
\newblock \emph{Phys. Rev. A}, 71\penalty0 (1):\penalty0 012307, 2005.
\newblock \doi{10.1103/PhysRevA.71.012307}.

\bibitem[Preskill(2018)]{preskill2018quantum}
John Preskill.
\newblock Quantum computing in the {NISQ} era and beyond.
\newblock \emph{Quantum}, 2:\penalty0 79, 2018.
\newblock \doi{10.22331/q-2018-08-06-79}.

\bibitem[Rigetti and Devoret(2010)]{rigetti2010fully}
Chad Rigetti and Michel Devoret.
\newblock Fully microwave-tunable universal gates in superconducting qubits with linear couplings and fixed transition frequencies.
\newblock \emph{Phys. Rev. B}, 81\penalty0 (13):\penalty0 134507, 2010.
\newblock \doi{10.1103/PhysRevB.81.134507}.

\bibitem[Roushan et~al.(2017)Roushan, Neill, Tangpanitanon, Bastidas, Megrant, Barends, Chen, Chen, Chiaro, Dunsworth, et~al.]{roushan2017spectroscopic}
Pedram Roushan, Charles Neill, J~Tangpanitanon, Victor~M Bastidas, A~Megrant, Rami Barends, Yu~Chen, Z~Chen, B~Chiaro, A~Dunsworth, et~al.
\newblock Spectroscopic signatures of localization with interacting photons in superconducting qubits.
\newblock \emph{Science}, 358\penalty0 (6367):\penalty0 1175--1179, 2017.
\newblock \doi{10.1126/science.aao1401}.

\bibitem[Sheldon et~al.(2016)Sheldon, Magesan, Chow, and Gambetta]{sheldon2016procedure}
Sarah Sheldon, Easwar Magesan, Jerry~M Chow, and Jay~M Gambetta.
\newblock Procedure for systematically tuning up cross-talk in the cross-resonance gate.
\newblock \emph{Phys. Rev. A}, 93\penalty0 (6):\penalty0 060302(R), 2016.
\newblock \doi{10.1103/PhysRevA.93.060302}.

\bibitem[Smith et~al.(2019)Smith, Kim, Pollmann, and Knolle]{smith2019simulating}
Adam Smith, MS~Kim, Frank Pollmann, and Johannes Knolle.
\newblock Simulating quantum many-body dynamics on a current digital quantum computer.
\newblock \emph{npj Quantum Inf.}, 5\penalty0 (1):\penalty0 1--13, 2019.
\newblock \doi{10.1038/s41534-019-0217-0}.

\bibitem[Tripathi et~al.(2019)Tripathi, Khezri, and Korotkov]{tripathi2019operation}
Vinay Tripathi, Mostafa Khezri, and Alexander~N Korotkov.
\newblock Operation and intrinsic error budget of a two-qubit cross-resonance gate.
\newblock \emph{Phys. Rev. A}, 100\penalty0 (1):\penalty0 012301, 2019.
\newblock \doi{10.1103/PhysRevA.100.012301}.

\bibitem[Trotter(1959)]{trotter1959product}
Hale~F Trotter.
\newblock On the product of semi-groups of operators.
\newblock \emph{Proceedings of the American Mathematical Society}, 10\penalty0 (4):\penalty0 545--551, 1959.
\newblock \doi{10.2307/2033649}.

\bibitem[Vovrosh and Knolle(2021)]{vovrosh2021confinement}
Joseph Vovrosh and Johannes Knolle.
\newblock Confinement and entanglement dynamics on a digital quantum computer.
\newblock \emph{Sci. Rep.}, 11\penalty0 (1):\penalty0 1--8, 2021.
\newblock \doi{10.1038/s41598-021-90849-5}.

\bibitem[Vovrosh et~al.(2021)Vovrosh, Khosla, Greenaway, Self, Kim, and Knolle]{vovrosh2021simple}
Joseph Vovrosh, Kiran~E Khosla, Sean Greenaway, Christopher Self, Myungshik~S Kim, and Johannes Knolle.
\newblock Simple mitigation of global depolarizing errors in quantum simulations.
\newblock \emph{Phys. Rev. E}, 104\penalty0 (3):\penalty0 035309, 2021.
\newblock \doi{10.1103/PhysRevE.104.035309}.

\bibitem[Wang et~al.(2015)Wang, Deng, and Duan]{wang2015hamiltonian}
Sheng-Tao Wang, Dong-Ling Deng, and Lu-Ming Duan.
\newblock Hamiltonian tomography for quantum many-body systems with arbitrary couplings.
\newblock \emph{New J. Phys.}, 17\penalty0 (9):\penalty0 093017, 2015.
\newblock \doi{10.1088/1367-2630/17/9/093017}.

\bibitem[Wilkinson and Hartmann(2020)]{wilkinson2020superconducting}
Samuel~A Wilkinson and Michael~J Hartmann.
\newblock Superconducting quantum many-body circuits for quantum simulation and computing.
\newblock \emph{Appl. Phys. Lett.}, 116\penalty0 (23):\penalty0 230501, 2020.
\newblock \doi{10.1063/5.0008202}.

\bibitem[You et~al.(2022)You, Huang, Alyanak, Romanenko, Grassellino, and Zhu]{you2022qcb}
Xinyuan You, Ziwen Huang, Ugur Alyanak, Alexander Romanenko, Anna Grassellino, and Shaojiang Zhu.
\newblock {Stabilizing and Improving Qubit Coherence by Engineering the Noise Spectrum of Two-Level Systems}.
\newblock \emph{Phys. Rev. Applied}, 18\penalty0 (4):\penalty0 044026, 2022.
\newblock \doi{10.1103/PhysRevApplied.18.044026}.

\bibitem[Zhu et~al.(2022)Zhu, Sun, Gong, Chen, Zhang, Wu, Ye, Zha, Li, Guo, et~al.]{zhu2022observation}
Qingling Zhu, Zheng-Hang Sun, Ming Gong, Fusheng Chen, Yu-Ran Zhang, Yulin Wu, Yangsen Ye, Chen Zha, Shaowei Li, Shaojun Guo, et~al.
\newblock Observation of thermalization and information scrambling in a superconducting quantum processor.
\newblock \emph{Phys. Rev. Lett.}, 128\penalty0 (16):\penalty0 160502, 2022.
\newblock \doi{10.1103/PhysRevLett.128.160502}.

\end{thebibliography}

\end{document}